\documentclass[twocolumn,showpacs,preprintnumbers,amsmath,amssymb]{revtex4}

\usepackage{graphicx}
\usepackage{dcolumn}
\usepackage{bm}
\begin{document}
\title
{Optimizing synchronizability of networks }
\author{Bing Wang$^{1,2}$}
\email{bingbignmath@yahoo.com.cn}
\author{Huanwen Tang$^1$}
\author{Tao Zhou$^3$}
\email{zhutou@ustc.edu}
\author{Zhilong Xiu$^4$}
\affiliation {$^1$ Department of Applied Mathematics, Dalian
University of Technology, Dalian Liaoning, 116024, P. R. China }
\affiliation {$^2$ Department of Physics, BK21 Physics Research
Division, and Institute of Basic Science, Sungkyunkwan University,
Suwon 440-746, Korea} \affiliation{$^3$ Department of Modern
Physics and Nonlinear Science Center, University of Science and
Technology of China, Hefei Anhui, 230026, P. R. China}
\affiliation{$^4$ School of Environmental and Biological Science
and Technology, Dalian University of Technology, Dalian Liaoning,
116024, P. R. China}
\begin{abstract}
In this paper, we investigate the factors that affect the
synchronization of coupled oscillators on networks. By using the
edge-intercrossing method, we keep the degree distribution unchanged
to see other statistical properties' effects on network's
synchronizability. By optimizing the eigenratio $R$ of the coupling
matrix with \textit{Memory Tabu Search} (MTS), we observe that a
network with lower degree of clustering, without modular structure
and displaying disassortative connecting pattern may be easy to
synchronize. Moreover, the optimal network contains fewer small-size
loops. The optimization process on scale-free network strongly
suggests that the heterogeneity plays the main role in determining
the synchronizability.
\end{abstract}
 \maketitle

\section{Introduction}
Many real networked systems exhibit some common characteristics,
including small-world effect \cite{Watts98} and scale-free property
\cite{Barabasi99}. The network structure has significant impact on
the dynamical processes taking place on it
\cite{Reviews1,Reviews2,Reviews3,Reviews4}. Recently, a particular
issue is concerned the synchronization of individuals with coupling
dynamics located at each node of a given network. It has been shown
that the ability of a network to synchronize is generally improved
in both small-world networks and scale-free networks as compared to
regular graphs \cite{Lago-Fernndez00,Gade00,Wang02}. However, so
far, a completely clear picture about the relationship between
structure and synchronizability is lacking.

How does the network topology affects its synchronizability? Some
previous works have shown that the average network distance $\langle
{d}\rangle$ is one of the key factors that affect the network
synchronizability. The smaller $\langle {d}\rangle$ will lead to
better synchronizability \cite{Barahona02,Lind04,ZhouT06}, while
other researchers found that the heterogeneity of degree can
strongly influence a network's synchronizability, that is, a more
homogeneous network will show better synchronizability compared to
the heterogeneous one, though the average distance of a homogeneous
network is longer \cite{Nishikawa03,Hong04,Zhao05}. In addition,
some recent works demonstrate that the disassortative networks
synchronize easier than assortative ones \cite{Bernardo05}, and the
increasing clustering will hinder the global synchronization
\cite{McGrawWu05,Wu06}.

\begin{figure}
\begin{center}
\scalebox{0.5}[0.5]{\includegraphics{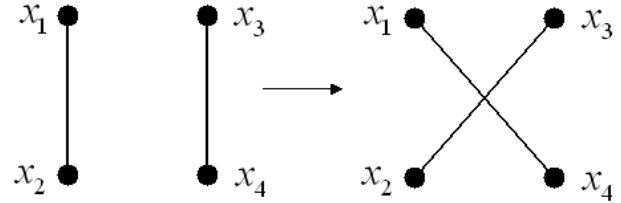}} \caption{The
sketch maps of edge-intercrossing operation.}
\end{center}
\end{figure}

The majority of previous works are based on the simulation results
allowing tuning only one or a very few topological measures (see
also some recently proposed analytical approaches
\cite{Hasegawa04,Hasegawa05,Munoz03,Munoz05}). If all the
topological properties can simultaneously vary, what will happen?
Should a network with better synchronizability have some particular
properties? Besides the simulated and analytical approaches, a
potential way to investigate the relation between structural and
dynamical properties is to track the optimization process, which
will lead to some networks with specific dynamical characters
\cite{Paul2004,Wang2006,Liu2005}. Concerning with network
synchronizability, one pioneer work \cite{Donetti05}, based on a
modified simulated annealing algorithm, suggested that the network
having best synchronizability should be extremely homogeneous. In
Donetti-Hurtado-Mu\~{n}oz approach \cite{Donetti05}, the rewiring
operation is used, thus for different initial configurations, the
optimization process will lead to the same optimal result, named
\emph{Entangled Network}. In this paper, as a complementary work, we
would like to see the optimal results of keeping the degree of each
node unchanged. With this idea, we take the network
synchronizability as the object and change the network structure
with the constraint condition that the degree of each node could not
be changed, which can be realized by edge-intercrossing operation
\cite{Maslov02,Kim04,ZhaoM06}. As shown in Fig. 1, the procedure of
edge-intercrossing operation is as follows: (1) Randomly pick two
existing edges $e_1=x_1x_2$ and $e_2=x_3x_4$, such that $x_1\neq
x_2\neq x_3\neq x_4$ and there is no edge between $x_1$ and $x_4$ as
well as $x_2$ and $x_3$. (2) Interchange these two edges, that is,
connect $x_1$ and $x_4$ as well as $x_2$ and $x_3$, and remove the
edges $e_1$ and $e_2$ simultaneously.

This article is organized as follows: The concept of
synchronizability and the optimization algorithm are briefly
introduced in Sec. II and Sec. III, respectively. Then, in Sec.
IV, we will give out the main simulation results for both
homogenous and heterogenous networks. Finally, in Sec. V, we will
summarize this work.

\section{Synchronizability}
Network synchronizability can be quantified by the eigenratio of
the Laplacian matrix \textbf{L}. Consider a network of \emph{N}
identical systems with symmetric coupling between oscillators. The
equations of motion for the system are:
\begin{equation}
\dot{\textbf{x}}_i=\textbf{F}(\textbf{x}_i)-\sigma\sum_{j=1}^N\textbf{L}_{ij}\textbf{H}(\textbf{x}_j),
\end{equation}
where $\dot{\textbf{x}}_i=\textbf{F}(\textbf{x}_i)$ governs the
dynamics of individual oscillator, \textbf{H(x)} is the output
function and $\sigma$ is the coupling strength. The $N\times{N}$
Laplacian matrix \textbf{L} is given by
\begin{equation}
    \textbf{L}_{ij}=\left\{
    \begin{array}{cc}
    k_i   &\mbox{for $i=j$}\\
     -1    &\mbox{for $j\in\Lambda_i$}\texttt{   },  \\
     0    &\mbox{otherwise}
    \end{array}
    \right.
\end{equation}
where $\Lambda_i$ denotes the set of $i$'s neighbors. All the
eigenvalues of Laplacian matrix \textbf{L} are positive reals and
the smallest eigenvalue $\lambda_{1}$ is zero for the rows of
\textbf{L} having zero sum. The eigenvalues are
$0=\lambda_{1}\leq{\lambda_{2}}\ldots\leq{\lambda_{N}}$. The ratio
of the maximum eigenvalue $\lambda_{N}$ to the smallest nonzero
eigenvalue $\lambda_{2}$ is widely used to measure the
synchronizability of the network \cite{Pecora1998,Pecora2005}. If
the eigenratio $R$ satisfies
$R=\frac{\lambda_{N}}{\lambda_{2}}<\beta$, where ${\beta}$ is a
constant depending on $\textbf{F}(x)$ and $\textbf{H}(x)$, then the
network is synchronizable. The eigenratio $R$ depends only on the
topology of interactions among oscillators. The impact of having a
particular coupling topology on the network's synchronizability is
represented by a single quantity
$R=\frac{\lambda_{N}}{\lambda_{2}}$: The smaller the eigenratio $R$,
the easier it is to synchronize the oscillators, and vice versa.
Having reduced the problem of optimizing the network
synchronizability to finding the smallest eigenratio $R$ of the
matrix \textbf{L}, we shall seek the network with approximately best
synchronizability by using edge-intercrossing operation.

\begin{center}
\begin{figure}
\scalebox{0.7}[0.7]{\includegraphics{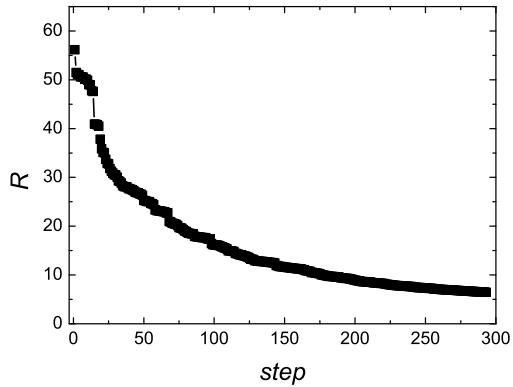}} \caption{The
eigenratio $R$ vs. iteration steps of a WS network with rewiring
probability $p=0.1$. The network size is $N=400$ and the average
degree is $\langle{k}\rangle=6$. Only the steps in which $R$ being
reduced are recorded.}
\end{figure}
\end{center}

\section{Algorithm}
To obtain the approximately minimal $R$, we combine the
intercrossing processes with a heuristic algorithm, named Memory
Tabu Search (MTS). The whole processes of the algorithm are (see
also Refs. \cite{Ji04,Glovera1989,Glovera1990} for the details about
MTS):

$step\ 1$: Generate an initial random graph $G_0$ with $N$ nodes,
$E$ edges following a previously given degree distribution. Set
$G^*=G_0$ and compute the eigenratio $R^*$ of the Laplacian matrix
of $G^*$. $G^*$ and $R^*$ are used to record the current optimal
configuration and the corresponding eigenratio, respectively.

$step\ 2$: Stop and output the current result $G^*$ and $R^*$ if a
prescribed terminal condition is satisfied. Otherwise, do intercross
two randomly selected edges of $G_k$ ($k$ denotes the time step, and
is set as zero initially), and denote by $G'$ and $R'$ the resulting
graph and its eigenratio.

$step\ 3$: Update the current optimal eigenratio $R^*$ and set the
current optimal graph as $G^*=G'$ if the eigenratio $R'$ satisfies
$R'<R^*$.

$step \ 4$: Update the new iteration resolution $G_{k+1}$. If
$R'<R_{G_k}$ or $G'$ does not satisfy the tabu condition, set
$G_{k+1}=G'$; else set $G_{k+1}=G_{k}$.

$step \ 5$: Update the tabu list. Let $G_{k+1}$ enter into the tabu
list. If $k\geq{L}$, where $L$ is the length of tabu list, delete
$G_{k+1-L}$ from the tabu list.

$step \ 6$: Return to $step\ 2$ and set $k=k+1$, where $k$ always
denotes the current time step.

The following condition is used to determine if a move is tabu:
$\frac{|R_{G_{k}}-R_{G'}|}{R_{G_{k}}}>\delta$, which is the
percentage improvement or destruction that will be accepted if the
new move is accepted. Thus, the new graph $G'$ at $step\ 2$ is
assumed tabu if the total change in the objective function is higher
than a percentage $\delta$. In this paper, $\delta$ is a random
number following uniform distribution between 0.50 and 0.75 and the
selection of $\delta$ will not affect the optimal result. The
terminal condition is that the present step is getting to the
predefined maximal iteration steps.

\begin{figure}
\begin{center}
\scalebox{0.7}[0.7]{\includegraphics{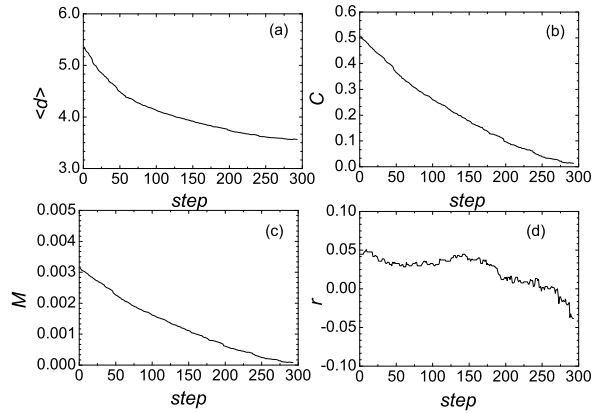}} \caption{Some
topological measures of the network vs. iteration steps. (a) The
average distance $\langle{d}\rangle$; (b) The average clustering
coefficient $C$; (c) Modularity $M$; (d) The Pearson parameter of
degree-degree coefficient $r$. As the same as shown in Fig. 2, the
initial network is a WS network with $N=400$, $\langle{k}\rangle=6$
and $p=0.1$.}
\end{center}
\end{figure}

\section{Simulation results}
In this paper, four different kinds of extensively studied networks
are used for simulation. These are scale-free networks proposed by
Barab\'{a}si and Albert (BA) \cite{Barabasi99}, small-world networks
proposed by Watts and Strogatz (WS) \cite{Watts98}, random networks
proposed by  Erd\"{o}s and R\'{e}nyi (ER) \cite{Erdos59}, and the
regular networks. The former one is heterogenous, while the latter
three are homogenous. The simulation results of all the homogenous
networks are almost the same, so, in this article, only the results
of WS networks are shown.

Fig. 2 reports the eigenratio $R$ vs. the iteration steps of a WS
network with rewiring probability $p=0.1$. The tendencies of some
other topological measures are shown in Fig. 3, these includes the
average distance $\langle {d}\rangle$, the clustering coefficient
$C$, the assortativity $r$ \cite{Newman02}, and the modularity $M$
\cite{Pimm1980,Newman01a,Newman01b}. Based on Pimm's work
\cite{Pimm1980}, $M$ is defined as
\begin{equation}
M=\frac{1}{N(N-1)}\sum_{i=1}^{N}\sum_{j=1}^{N}{S_{ij}},
\end{equation}
where $S_{ij}$ is the number of common neighbors of nodes $i$ and
$j$ divided by their total number of neighbors.

One trivial result is that the networks with better
synchronizability are of shorter average distance. Besides, we
observe that the degree correlation coefficient decreases from
positive reals to negative ones, that is, the disassortative
networks are, in general, easier to synchronize. And, in the whole
process, the average clustering coefficient reduces monotonically.
The lower clustering and disassortative pattern of the optimal
network are consistent with some previous results
\cite{Bernardo05,McGrawWu05,Wu06}. More interestingly, the
modularity decreases too, which indicates that a network with strong
modular structures may be more difficult to synchronize, which is
also in accordance with some recent results \cite{Zhou2006}.

\begin{figure}
\begin{center}
\scalebox{0.7}[0.7]{\includegraphics{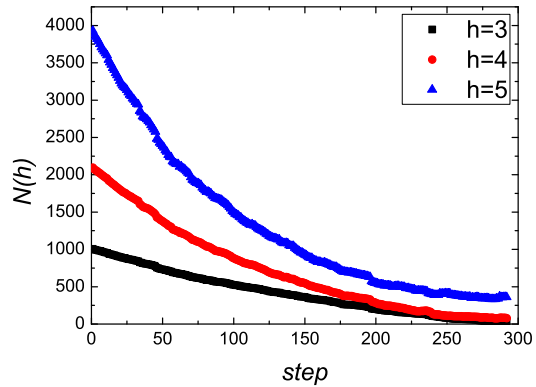}} \caption{(Color
online) Number of $h$-loops, $N_h$, in the network vs. iteration
steps. The value of $N_{h}$ in step zero corresponds to the case of
initial network, which is a WS network with $N=400$,
$\langle{k}\rangle=6$ and $p=0.1$.}
\end{center}
\end{figure}

It has been found that many biological and technological networks
contain motifs, that is, some specific subgraphs appearing much more
frequently than that observed in random graphs with the same degree
sequence \cite{Shen02,Milo02}. Loop is one of the simplest but most
significant subgraphs, for it accounts for the multiplicity of paths
between any two nodes \cite{Bianconi05}. Here we will show the
change of loop structure in the optimization process. Due to the
limitation of computational ability, we only calculate the number of
loops from size 3 to 5. From Fig. 4, one can observe that the number
of loops drops drastically during the optimization process, which
indicates that the dense loops may hinder the global
synchronization. As shown in Fig. 5, in WS networks, the number of
loops will increase linearly as the increasing of network size,
however, the number of loops in the optimal networks is stable and
much smaller than that in the initial networks.

All these results indeed state that the optimal networks belong to
a class of networks in which there are few number of loops, which
may be different from the majority of real biological and
technological networks. This phenomenon may deserve deep
explorations in local structures of both artificial and real
networks.

\begin{figure}
\begin{center}
\scalebox{0.7}[0.6]{\includegraphics{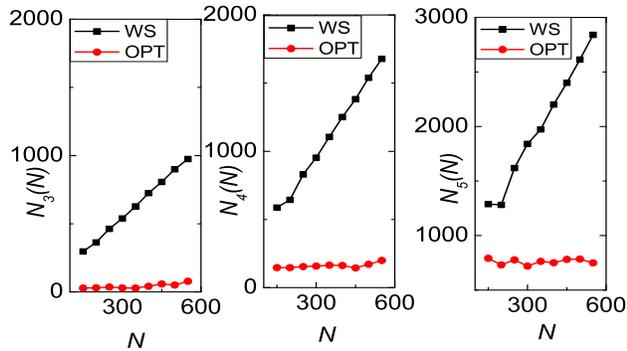}} \caption{(Color
online) Number of $h$-loops, $N_{h}{(N)}$, vs. $N$ in WS networks
(WS, black squares) and the corresponding optimal networks (OPT, red
circles), respectively. The network size grows from $N=150$ up to
$N=550$. In all those simulations, the rewiring probability and
average degree are $p=0.1$ and $\langle{k}\rangle=6$. The three
panels from left to right are for $h=3$, $h=4$ and $h=5$,
respectively.}
\end{center}
\end{figure}

As shown before, this algorithm is highly effective for homogenous
networks. For example, the eigenratio of WS networks can be reduced
to about $15\%$ of the original value (see Fig. 2). Since the degree
distributions of many real networks are heterogenous, next, we will
check if this algorithm is also effective for heterogenous networks.
As shown in Fig. 6, this algorithm can also enhance the
synchronizability of scale-free networks, however, the enhancement
is very small compared with the case of homogenous networks. In the
optimization process of BA network, the tendencies of average
distance, clustering coefficient, modularity and assortativity are
almost the same as those for WS network (not shown), but with larger
fluctuation and smaller decrement. Different from the case of
homogenous networks, as the increasing of network size, the number
of loops in both the optimal network and original BA network
increases (see Fig. 7). The number of loops in the optimal network
is a little bit smaller than that in the original BA network.

\begin{figure}
\begin{center}
\scalebox{0.7}[0.7]{\includegraphics{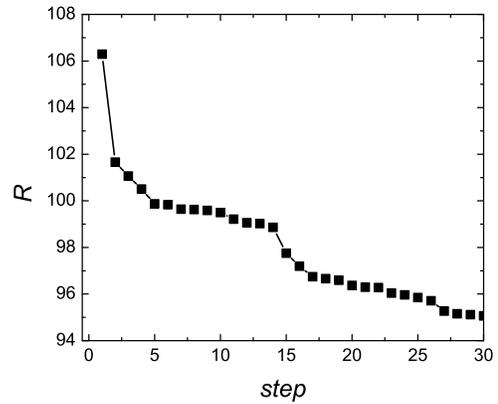}} \caption{The
eigenratio $R$ vs. iteration steps of a BA network with $N=500$ and
$\langle{k}\rangle=4$. Only the steps in which $R$ being reduced are
recorded.}
\end{center}
\end{figure}

\begin{figure}
\begin{center}
\scalebox{0.7}[0.6]{\includegraphics{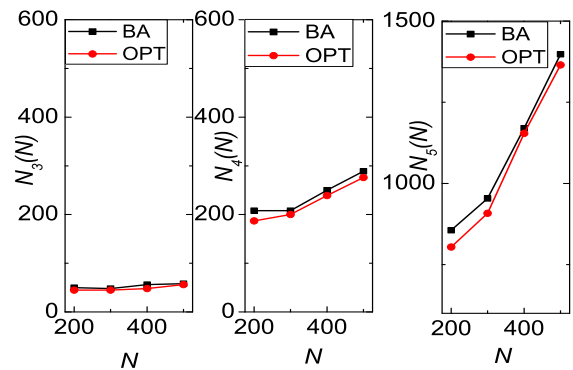}} \caption{(Color
online) Number of $h$-loops, $N_{h}{(N)}$, vs. $N$ in BA networks
(BA, black squares) and the corresponding optimal networks (OPT, red
circles), respectively. The network size grows from $N=200$ up to
$N=500$. In all those simulations, the average degree is
$\langle{k}\rangle=4$. The three panels from left to right are for
$h=3$, $h=4$ and $h=5$, respectively.}
\end{center}
\end{figure}

\section{Conclusion and Discussion}
In summary, a heuristic algorithm, memory tabu search, is used to
optimize the network synchronizability by changing the connection
pattern between different pairs of nodes while keeping the degree of
each node unchanged. For both the homogenous and heterogenous
networks, the change of some topological measures is recorded during
the optimization process, which strongly suggests that a network
with shorter average distance, lower clustering, negative
degree-degree correlation and weaker modular structure may be easier
to synchronize. In fact, these results has been reported
respectively by some previous works. However, most of those works
are based on some specific network models and can only reveal the
relation between synchronizability and one or two structural
properties, while the present optimization algorithm can
simultaneously lay out the whole picture. Here, we would like to
emphasize that, besides the traditional simulated and analytical
approaches, to track the optimization process is a powerful tool of
analysis on the relation between structural and dynamic properties
of networks.

In addition, we investigated the change of loop structure in the
optimization process, and found that the number of loops will
decrease as the increasing of synchronizability. That is to say, the
networks with fewer loops will be easier to synchronize. More
interestingly, in the original WS networks, the number of loops will
increase linearly as the increasing of network size, while in the
optimal networks, this number keeps stable. This novel phenomenon
deserves in-depth exploration. Since each node can only impact its
neighbors (see Eq. (1)), if there is one path, of length $l$,
between node $i$ and $j$, then, along this path, it takes $l$ steps
transferring the synchronization signal from $i$ to $j$ (or from $j$
to $i$). Therefore, if there are so many paths of different lengths
between node $i$ and $j$, the synchronization signal of $i$ at a
given time will arrive on $j$ along different paths at different
time, which may disturb each other. It may be the reason why dense
loops will hinder the global synchronization. An extreme case is
that for directed networks, the one with highest synchronizability
(i.e. with eigenratio $R$ being equal to 1) is a tree structure
without any loops \cite{Nishikawa2006}. Even if adding one loop of
length 2 \cite{ex1}, the eigenratio will be doubled
\cite{Chavez2005,Zhao2006}. However, the conclusion in Ref.
\cite{Nishikawa2006} may be not universal. Actually, in
synchronization system consisted of non-identical oscillators, a
counterexample against Ref. \cite{Nishikawa2006} is reported very
recently \cite{Um2007}. We believe this work will be helpful for the
in-depth understanding about the role of loops in network
synchronization.

Many previous works focus on synchronization on heterogenous
networks, especially scale-free networks. A common cognition is
that the heterogeneity will hinder the global synchronization.
However, since most of those works are based on some specific
models, it is not clear whether the homogeneity is a necessary
condition for strong synchronizability. Note that, for a fixed
degree sequence the present algorithm with edge-intercrossing
operation is ergodic, thus the less efficiency of the present
algorithm on scale-free networks indicate that the homogeneity
should be a necessary condition. That is to say, within the
framework of synchronization of identical oscillators with uniform
coupling mode, the heterogenous network is hard to synchronize in
despite of its idiographic structure.

\begin{acknowledgements}
The authors thank the guidance of Huanwen
Tang, and wish to recall the memory of Prof. Tang through this
paper. We also acknowledge W. X. Wang and C. H. Guo for their
helpful comments and suggestions. This work is supported by the
NNSFC under Grant Nos. 10571018 and 10635040.
\end{acknowledgements}


\begin{thebibliography}{Watts98}
\bibitem{ex1} In directed networks, two reverse edges between $i$
and $j$ can be considered as a loop of length two as $i\rightarrow
j \rightarrow i$.
\end{thebibliography}
\end{document}